\documentclass{article}
\usepackage{amsmath}
\usepackage{amssymb}
\usepackage{graphicx}
\usepackage{float}
\usepackage{authblk}
\usepackage{cite}
\usepackage[left=3cm, right=2cm, top=2cm, bottom=2cm]{geometry}
\usepackage{hyperref}
\usepackage{fancyhdr}
\graphicspath{{fig/}}
\usepackage{caption}
\begin{document}

\title{\textbf{Fallacies on pairing symmetry and intrinsic electronic Raman spectrum in high-$T_c$ cuprate superconductors}}

\author [1,2,\footnote{\href{mailto:htkim@etri.re.kr}{htkim@etri.re.kr, htkim580711@gmail.com}}]{Hyun-Tak Kim}

\font\myfont=cmr12 at 10pt
\affil[1]{\myfont Metal-Insulator Transition Lab., Electronics and Telecommunications Research Institute, Daejeon 34129, South Korea}
\affil[2]{\myfont School of Advanced Device Technology, University of Science and Technology, Daejeon 34113, South Korea}
\date{}

\maketitle

\vspace*{-1cm}

\begin{abstract}
\font\myfont=cmr12 at 12pt
\noindent{\myfont Certain significant fallacies are involved in discussions of the high-$T_c$ mechanism unsolved for over 30 years in cuprate superconductors. These fallacies are explored with the aim of unravelling this mechanism. Moreover, using polarised electronic Raman scattering in inhomogeneous underdoped cuprate superconductors, the intrinsic nonlinear Raman spectrum is obtained by subtracting the pseudogap characteristic of a nonlinear from the linear Raman spectrum measured in the $B_{2g}$ mode of the node area below the critical temperature. The intrinsic nonlinear behaviour implies the existence of the nodal superconducting gap denying $d_{x^{2}-y^{2}}$-wave pairing symmetry. An origin of the nodal superconducting gap is discussed.}

\end{abstract}


\section{Fallacies in the high-$T_c$ mechanism in cuprate superconductors} 
\font\myfont=cmr12 at 12pt

\myfont Since the discovery of high-$T_c$ cuprate superconductors in 1986, many researchers have attempted to reveal the high-$T_c$ mechanism.\cite{shen1993anomalously,kim2007comments,kaminski2000quasiparticles,okawa2009superconducting,kim2018analysis,tsuei2000pairing,welp1994magneto,kim1996paramagnetic,klemm1997we,tsuei1996pairing,devereaux1994electronic,kruchinin2011modern,kruchinin2014physics,qazilbash2005evolution,chen1997electronic,blanc2010loss,
hiramachi2007polarization,zhang2003new,kim2017high,klemm2012layered,yoshida2011pseudogap,kaminski2015pairing,razzoli2010fermi,rice2011phenomenological,shen2005nodal,yagi2005photoemission,sassa2011arpes,kim2019interplay,kim2002analyzing,di2017disorder,kim2016impurity,kim2016photoheat} The number of published papers on this topic is over 200,000; many scientific experimental systems have been carefully developed, many types of high-quality single crystals have been grown, and, in particular, many theorists have been focused on developing $d_{x^{2}-y^{2}}$ (\textit{d})-wave (Fig. 1a) theories with a large anti-nodal superconducting gap proportional to $T_c$. Nevertheless, the mechanism of high-$T_c$ superconductivity is still shrouded in mystery, and this has become a historic problem in the field of condensed matter physics. What are the obstacles we need to overcome? We notice that there are significant fallacies in analyses of experimental data, and that corrections of these can be used as sources for identifying the mechanism. Here, we investigate these important fallacies. 

Firstly, we take a look at the analysis of the data measured by low-resolution angle-resolved photoemission spectroscopy (ARPES) system.\cite{shen1993anomalously,kim2007comments} The analysis shows that the behaviour of the ARPES spectrum, measured below $T_c$ at the node in an underdoped $Bi_2$$Sr_2$$Ca$$Cu_2$$O_{8+\delta}$ crystal, is similar to that above $T_c$ (B curves in Fig. 1b). This has been accepted by many researchers as evidence of \textit{d}-wave symmetry for the pairing symmetry of superconducting carriers, since \textit{d}-wave does not have a superconducting node gap but contains carriers at the nodes below and above $T_c$, and has a superconducting gap at the antinode. Additionally, a pseudogap near -0.1 eV in curve A at 20 K disappears at curve B of node (dot line) (Fig. 1b), which indicates that the pseudogap phase is \textit{d}-wave. After this research, a high-resolution ARPES system was developed, in which the measured data \cite{kaminski2000quasiparticles} showed a sharp clear nodal superconducting gap without broadening of no-gap characteristic (i.e. metallic characteristic) below $T_c$ (Fig. 1c),\cite{kim2007comments} unlike in the previous work.\cite{shen1993anomalously} Note that an ARPES spectrum (or peak) in metal is not observed, because it is given by a product of the imaginary part of the one-particle Green function and the Fermi distribution function. Moreover, in an optimally doped $Y$$Ba_2$$Cu_3$$O_{7-\delta}$ crystal, a nodal superconducting gap was clearly observed by laser ARPES (Fig. 1d).\cite{okawa2009superconducting} The nodal superconducting gaps are evidence refuting \textit{d}-wave symmetry and possible evidence for \textit{s}-wave symmetry.\cite{kim2018analysis} 

\begin{figure}[H]
\centering
\includegraphics [scale=0.72]{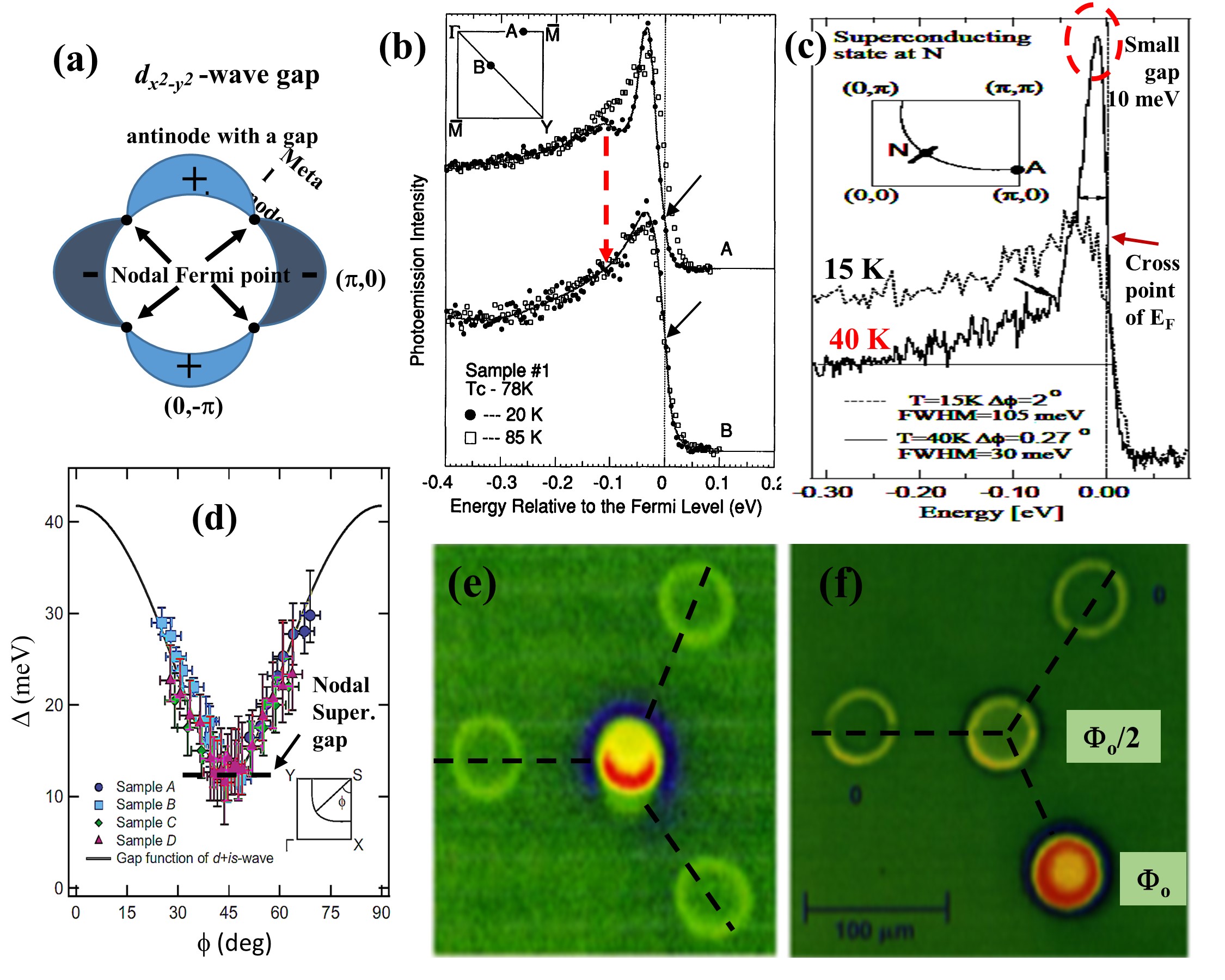}
\font\myfont=cmr12 at 11pt
\caption {\myfont \textbf {Previous data on pairing symmetry.} (a) The definition of the $d_{x^2-y^2}$ wave gap. (b) ARPES data measured in an underdoped $Bi_2$$Sr_2$$Ca$$Cu_2$$O_{8+\delta}$ crystal. The shapes of the curves measured at B (node) above and below $T_c$ are similar. In particular, the superconducting gap is not seen at B below $T_c$, which has been put forward as evidence of \textit{d}-wave symmetry (taken from Shen et al.\cite{shen1993anomalously}). (c) ARPES data measured in a high-resolution system in an optimally doped $Bi_2$$Sr_2$$Ca$$Cu_2$$O_{8+\delta}$ crystal, showing the obvious superconducting node gap at 40K (taken from Kaminski et al.\cite{kaminski2000quasiparticles}). Data at 15K was measured in a low resolution system. (d) A nodal superconducting gap, measured in an optimally doped $Y$$Ba_2$$Cu_3$$O_{7-\delta}$ crystal by a laser ARPES, was clearly seen (taken from Okawa et al.\cite{okawa2009superconducting}). (e) Top view of an image of the half-flux quantum measured by a SQUID microscope for a superconducting ring made from an underdoped $Y$$Ba_2$$Cu_3$$O_{7-\delta}$ film on the tricrystal point of a substrate (taken from Tsuei et al.\cite{tsuei2000pairing}). (f) A flux image measured in a superconducting ring for an overdoped $Tl_2$$Ba_2$$Cu$$O_{6+\delta}$ film (taken from Tsuei et al.\cite{tsuei1996pairing}); in the case of $\Phi_0$, the colour tone is uniform, which is evidence of textit{s}-wave symmetry. The dull colour image on the tricrystal point represents anisotropy.}
\end{figure}

Secondly, the half-flux quantum in a Josephson-$\pi$ junction with the Josephson tunnel effect (circulation of Josephson supercurrent), theoretically suggested as evidence of \textit{d}-wave pairing symmetry, was measured in an underdoped $Y$$Ba_2$$Cu_3$$O_{7-\delta}$ crystal ring on a tricrystal substrate in a very weak magnetic field, and was imaged using scanning SQUID (superconducting quantum interference device) microscopy (Fig. 1e).\cite{tsuei2000pairing} This was credited by researchers as an experimental observation of \textit{d}-wave symmetry. However, the image was anisotropic. Here, we found a fatal error that a magnetic field was applied into the junction, although the Josephson tunnel effect occurs in the absence of a magnetic field. When a magnetic field is applied to underdoped crystals with an incomplete Fermi surface, part of the magnetic field (or flux) is trapped by the superconducting phase.\cite{welp1994magneto,kim1996paramagnetic,klemm1997we} Thus, the anisotropic image is evidence of both trapped flux and the absence of circulating supercurrent (Fig. 1e).\cite{kim2018analysis} Further, a measurement of the flux quantum for an overdoped $Tl_2$$Ba_2$$Cu$$O_{\beta+\delta}$ crystal without magnetic field showed a strong ring image with a flux quantum, $\Phi_0$ (Fig. 1f).\cite{tsuei1996pairing} This is regarded as evidence of \textit{s}-wave symmetry.

Thirdly, for underdoped crystals with an incomplete Fermi surface, many researchers believe that the superconducting gap below $T_c$ is formed at the antinode despite the absence of the superconducting gap at the antinode, although the Fermi arc near the node has been measured by ARPES above $T_c$.\cite{okawa2009superconducting} This is a misplaced preconception.

These fallacies arise from the inhomogeneity of the cuprate superconductors, composed of a non-superconducting \textit{d}-wave pseudogap phase and an \textit{s}-wave-like superconductor phase.\cite{okawa2009superconducting} When these fallacies are rectified, the high-$T_c$ mechanism can be clarified.

\section{Intrinsic electronic Raman spectrum and pairing symmetry}

The pairing symmetry of superconducting carriers in inhomogeneous cuprate superconductors, which explains the high-$T_c$ mechanism, is believed by many to be $d_{x^{2}-y^{2}}$-wave with the electronic structure of the superconducting gap at the antinode and no gap at the node. This has remained unproven for over 30 years. Since polarized electronic Raman scattering (ERS) measures the characteristics of the bulk rather than the surface, unlike photoemission spectroscopy, ERS data (or spectra) can generate more accurate information on the characteristics of superconductors. One theory of ERS is that when $d_{x^{2}-y^{2}}$-wave symmetry is assumed, $I$($\omega$) $\propto$ $\omega$, linear behaviour, in the $B_{2g}$ mode and $I$($\omega$) $\propto$ $\omega^3$ for the $B_{1g}$ mode are calculated below a wave number corresponding to an energy gap.\cite{devereaux1994electronic} ERS experiments show the linear behaviour in the $B_{2g}$ mode measured at the node area below $T_c$ for underdoped crystals (Figs. 2a, 2b, and 2d).\cite{qazilbash2005evolution,chen1997electronic,blanc2010loss} The $\omega^3$ behaviour was investigated in optimally doped crystals although the linear behaviour did not show at the $B_{2g}$ mode.\cite{qazilbash2005evolution}

Since the measured data from inhomogeneous crystals with at least two phases has an averaged effect of the two phases to the measurement region, it is very difficult to determine the intrinsic effect. It is therefore necessary to decompose the measured data and analyse the intrinsic characteristics. ERS data well decomposed in an electron-doped cuprate $Pr_{2-x}$$Ce_x$$Cu$$O_{4-\delta}$ superconductor\cite{qazilbash2005evolution} is reviewed here (Figs. 2c-2e), allowing us to demonstrate the intrinsic effect. The authors measured a low-temperature curve (blue) at 4 K in the superconducting state and a curve (pink) in the normal state (Figs. 2c-2e).\cite{qazilbash2005evolution} The normal state curve was decomposed into Drude (green curve) and incoherent (black curve) components. The Drude component arose from free carriers in the normal state, and the incoherent component was caused by the pseudogap phase. Fig. 2c shows an absence of the superconducting gap in the $B_{1g}$ mode of the antinode. Both the superconducting gap and the linear behaviour in the $B_{2g}$ mode are represented by the blue curve in Fig. 2d. It was suggested that the low-temperature curve (blue) for the $B_{2g}$ mode is attributable to the Drude component (green) in the normal state.\cite{qazilbash2005evolution} The energy gap in the $B_{2g}$ mode was interpreted to be the \textit{d}-wave (or non-monotonic \textit{d}-wave) superconducting gap near the node, rather than at the node.\cite{qazilbash2005evolution}

As a comment, it should be noted that, since the ERS data have two effects arising from the superconducting and pseudogap phases, the incoherent pseudogap component (black) should be subtracted from the curve (blue) measured at low temperature, in order to reveal the intrinsic superconducting curve. The effect of the incoherent component is explained as an example. The incoherent curve (black) measured for a crystal with optimal doping in Fig. 2(e) is much smaller than that shown in Fig. 2(d), due to the insulator-metal transition induced by electron doping. The Drude curve (green) and the superconducting curve (blue) in Fig. 2(e) are much larger. The intrinsic superconducting curve (shown in orange) without the incoherent component follows a polynomial function of $y = 2.36072+0.11427x-0.0068x^2+1.39131\times10^{-4}x^3$ where $x$ is the wave number. This indicates that the incoherent component has a strong influence on the ERS data.

\begin{figure}[H]
\centering
\includegraphics [scale=0.55]{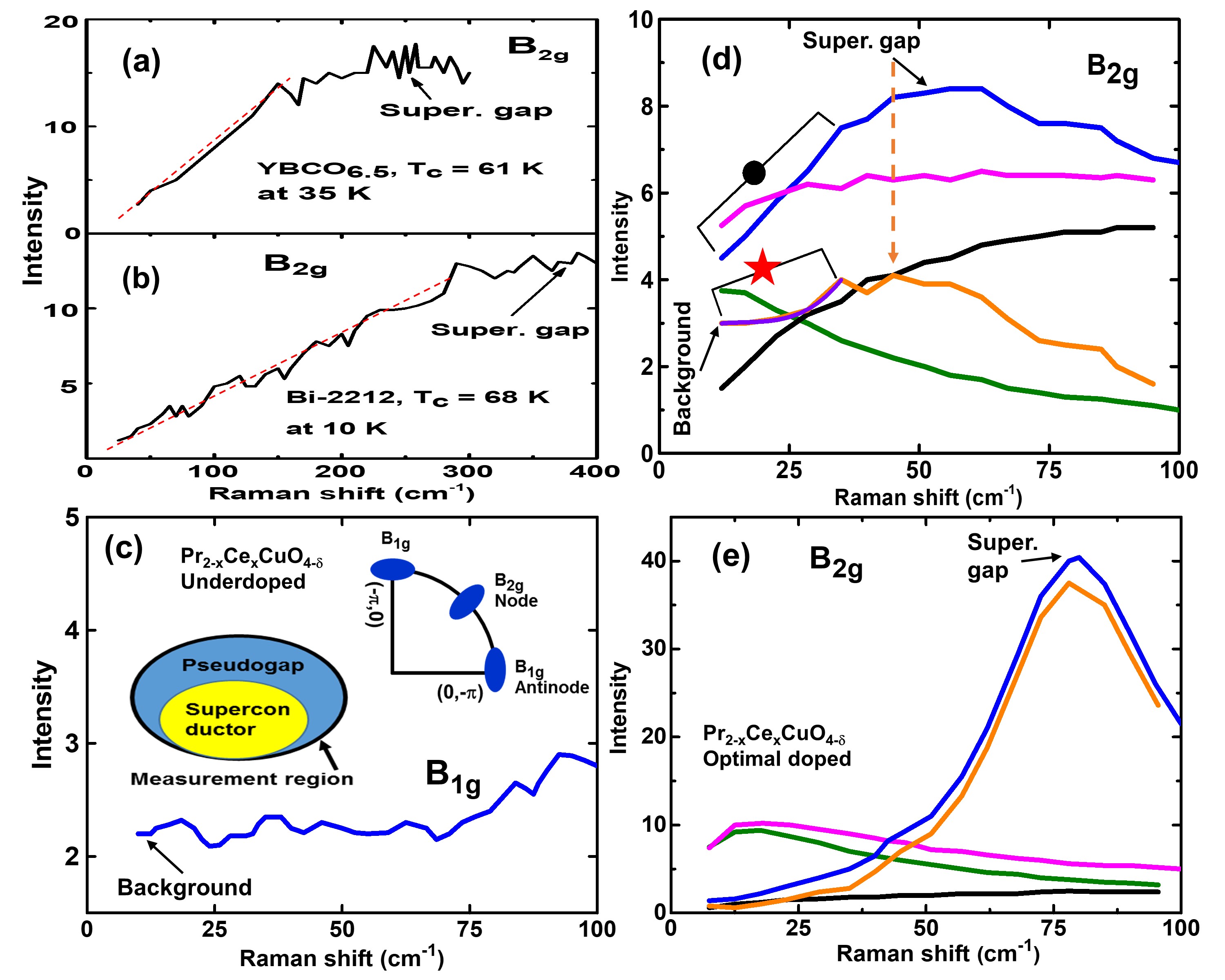}
\font\myfont=cmr12 at 11pt
\caption{\myfont \textbf{Analysis of the electronic Raman spectrum.} (a) The $B_{2g}$ mode measured at the node area using polarised ERS at 35 K for an underdoped $Y$$Ba_2$$Cu_3$$O_{6.5}$ single crystal with $T_c$ $\thickapprox$ 61K. Linear behaviour is shown below 150 $cm^{-1}$ and a superconducting gap is observed near 250 $cm^{-1}$. This curve was redrawn from the original data.\cite{chen1997electronic} (b) The $B_{2g}$ mode measured by ERS at 10 K for an underdoped $Bi_2$$Sr_2$$Ca$$Cu_2$$O_{8+\delta}$ single crystal with $T_c$ $\thickapprox$ 68 K. Linear behaviour is shown below 300 $cm^{-1}$ and a superconducting gap is exhibited near 380 $cm^{-1}$. This curve was redrawn from the original data.\cite{blanc2010loss} In both types of crystals, the \textit{d}-wave superconducting gap in the $B_{1g}$ mode of the antinode was not observed.\cite{chen1997electronic,blanc2010loss} (c) The $B_{1g}$ mode measured at the antinode area using ERS at 4 K in an underdoped $Pr_{2-x}$$Ce_x$$Cu$$O_{4-\delta}$ ($x=0.135$) single crystal with $T_c$ $\thickapprox$16.5 K. The \textit{d}-wave superconducting gap is not shown. The insets display the presence of two phases in the measurement region for the two-phase model and the measurement areas in \textit{k}-space of the $B_{1g}$ and $B_{2g}$ modes. (d) The $B_{2g}$ mode measured using ERS under the same conditions as for the crystal used in Fig. 1c. The pink curve was measured in the normal state; the green and black curves were decomposed by the pink curve. The blue curve measured at 4 K shows linear behaviour and a superconducting gap. The violet curve (red star) shows the fitting of the intrinsic superconducting curve (shown in orange) obtained by subtracting the incoherent part (black), giving the final result shown here. The small background is shown. (e) The $B_{2g}$ mode measured using ERS for an optimal doped $Pr_{2-x}$$Ce_x$$Cu$$O_{4-\delta}$ ($x=0.147$) single crystal with $T_c$ $\thickapprox$ 23.5 K. The pink curve was measured in the normal state; the green and black curves were decomposed by the pink curve. The blue curve measured at 4 K includes a superconducting gap. The intrinsic superconducting curve (orange) was obtained by subtracting the incoherent part (black). Figs. 2 (c, d, and e) were redrawn using data extracted from original data.\cite{qazilbash2005evolution}}
\end{figure}

In the case of the underdoped crystal (Fig. 2d), the intrinsic superconducting curve (orange) is obtained from subtracting the effect of the incoherent pseudogap phase (black) from the measured superconducting curve (blue). This fits the polynomial function $y = 0.7051-0.03343x-4.00818\times10^{-6}x^2+4.5587\times10^{-5}x^3$ (violet curve, red star), rather than a linear function, such as the intrinsic superconducting curve (orange) (Fig. 2e). This is a new result, because of which we suggest that the linear behaviors (Figs. 2a, 2b, and 2d) in the same context are also not intrinsic. Moreover, in previous researches, a similar curve on $B_{1g}$ for an overdoped $Ca$-$YBCO$ crystal\cite{hiramachi2007polarization} and a nonlinear curve at $B_{2g}$ in an optimally doped $La_{1.85}$$Sr_{0.15}$$Cu$$O_4$ crystal\cite{zhang2003new} were observed. Instead, the nonlinear behavior denying $d_{x^{2}-y^{2}}$-wave symmetry indicates that the super gaps (Figs. 2a, 2b, and 2d) are the superconducting gap like the Fermi-arc shape formed at and nearby the node, which is a starting point for the formation of the \textit{s}-wave gap.\cite{kim2018analysis,kim2017high} This may resolve the core problem of the high-$T_c$ mechanism. Further analyses of other experiments involving pairing symmetry are given.\cite{kim2017high,klemm2012layered}

\section{Origin of nodal superconducting gap}
Figures 1(c), 1(d), 2(a), 2(b), 2(d), and 2(e) show the nodal superconducting gap which should be absent in the $d_{x^{2}-y^{2}}$-wave symmetry. The forming mechanism of the gap is a core problem in the cuprate superconductor mechanism. An important phenomenon for the formation of the gap is Fermi arc with carriers at and near the node.\cite{kim2018analysis,qazilbash2005evolution, yoshida2011pseudogap,kaminski2015pairing,razzoli2010fermi,rice2011phenomenological,shen2005nodal,yagi2005photoemission,sassa2011arpes,kim2019interplay} The Fermi arc is caused from the \textit{d}-wave insulator-metal transition (IMT) which means that the insulating pseudogap phase changes into metal at the node.\cite{kim2007comments,kim2018analysis,kim2017high,yagi2005photoemission,kim2002analyzing} The IMT has been driven by disorder\cite{di2017disorder} or impurity.\cite{kim2017high,kim2016impurity,kim2016photoheat} Therefore, the nodal superconducting gap is formed below $T_c$ by carriers of Fermi arc generated by the \textit{d}-wave insulator-metal transition. This is schematically explained in Fig. 3.\cite{kim2018analysis,kim2017high}

\begin{figure}[H]
\centering
\includegraphics [scale=0.9]{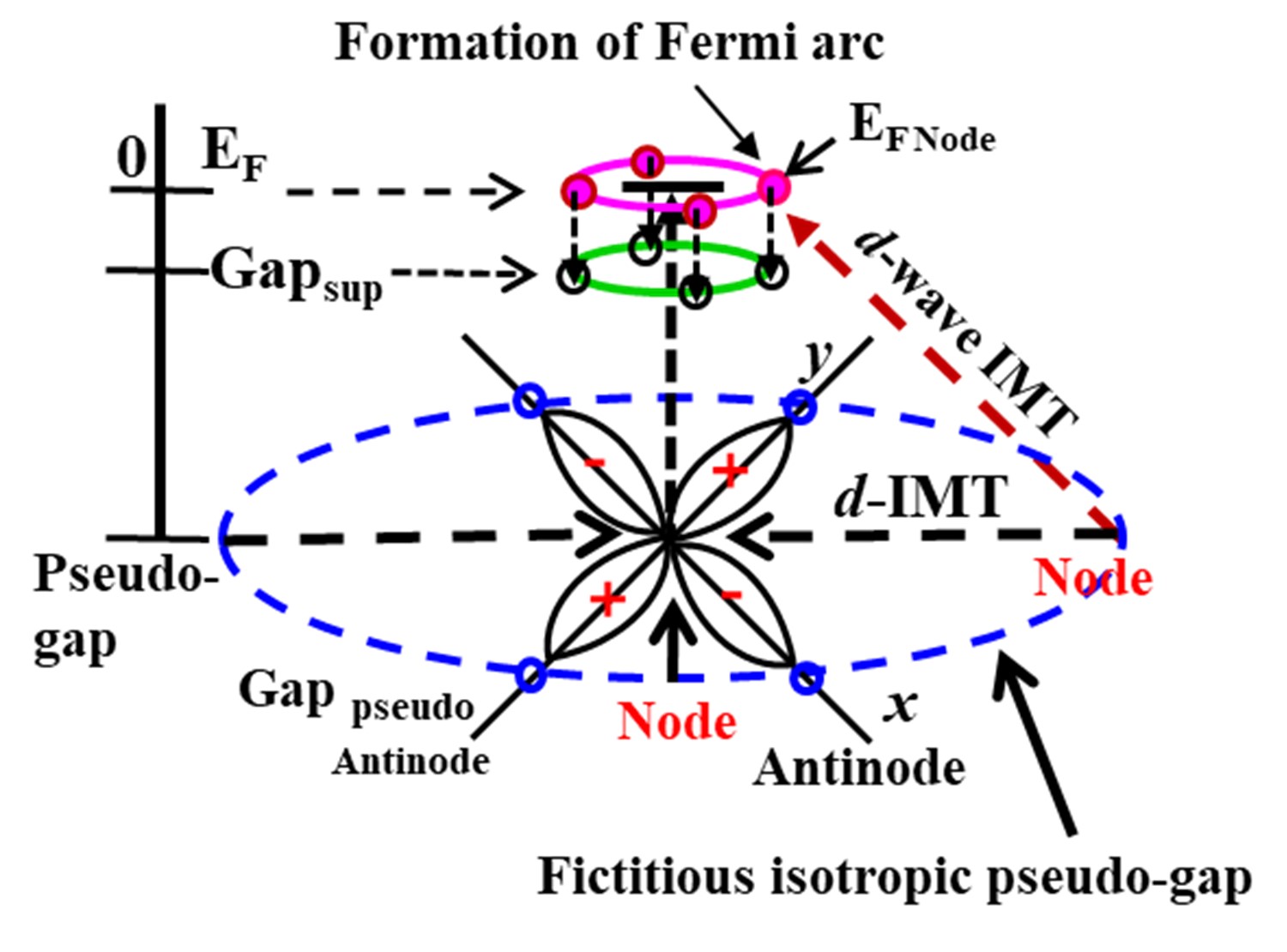}
\font\myfont=cmr12 at 11pt
\caption{\myfont \textbf{Forming mechanism of the nodal superconducting gaps for cuprate superconductors.\cite{kim2018analysis,kim2017high}} The presence of a fictitious isotropic pseudogap in a doped Mott insulator with a disorder or inhomogeneity is assumed. When excitation occurs by applying temperature, chemical doping, pressure etc to the doped Mott insulator, a transition from insulator to metal takes previously place at the node rather than the antinode. This is called the \textit{d}-wave insulator-metal transition. Then, the Fermi arc is formed and the nodal superconducting gap is formed below $T_c$. For optimal or over doping, the Fermi surface is completely formed. Finally, the complete \textit{s}-wave superconducting gap is formed.}
\end{figure}

\section{Conclusion}
The cause unsolved over 30 years of the cuprate superconductor mechanism was attributed to fallacies on analysis of experimental data. New analysis on the electronic Raman-spectrum data with linear characteristic, suggested as evidence of \textit{d}-wave symmetry, revealed a nonlinear behavior regarded as evidence of \textit{s}-wave symmetry. The superconducting gap is firstly formed at the node due to the \textit{d}-wave insulator-metal transition and develops from the node to the antinode with increasing doping.\cite{kim2018analysis,kim2017high} Finally, the \textit{s}-wave superconducting gap is made. Therefore, we maintain that pairing symmetry of Cooper pair in high-$T_c$ cuprate superconductors is \textit{s}-wave.

\vspace{1.0cm}

\noindent \textbf{Acknowledgments}

\noindent \small{We acknowledge M. M. Qazilbash, A. Kaminski, S. Shin, and M. Okawa for permissions using data, very valuable comments and proofreading for section 1 on the fallacies of pairing symmetry. We acknowledge Hyejin Lee so much for data extraction from original figures, and M. M. Qazilbash for very valuable comments, and G. Blumberg, C. Kendziora, R. Greene and W. Hardy for using their original data in section 2 on analysis of the intrinsic Raman scattering. This was supported by both the principal projects (19ZB1320) in ETRI and a MS and ICT project (2017-0-00830) on MIT.}


\begin{thebibliography} {}

\bibitem{shen1993anomalously} Shen, Z.-X. and Dessau, D.S. and Wells, B.O. and King, D.M. and Spicer, W.E. and Arko, A.J. and Marshall, D. and Lombardo, L.W. and Kapitulnik, A. and Dickinson, P. and others. Anomalously large gap anisotropy in the $a-b$ plane of ${Bi_2}$${Sr_2}$${Ca}$${Cu_2}$${O_{8+\delta}}$. \href{https://doi.org/10.1103/PhysRevLett.70.1553}{\emph{Phys. Rev. Lett.}} {\bf 70}, 10, 1553 (1993).

\bibitem{kim2007comments} Kim, H.-T. and Kim, B.-J. and Kang, K.-Y. Comments on identity of gap anisotropy and nodal constant Fermi velocity in ${Bi_2}$${Sr_2}$${Ca}$${Cu_2}$${O_{8+\delta}}$. \href{https://doi.org/10.1016/j.physc.2007.03.337}{\emph{Physica C: Superconductivity}} {\bf 460}, 943-945 (2007).

\bibitem{kaminski2000quasiparticles} Kaminski, A. and Mesot, J. and Fretwell, H. and Campuzano, J.C. and Norman, M.R. and Randeria, M. and Ding, H. and Sato, T. and Takahashi, T. and Mochiku, T. and others. Quasiparticles in the superconducting state of ${Bi_2}$${Sr_2}$${Ca}$${Cu_2}$${O_{8+\delta}}$. \href{https://doi.org/10.1103/PhysRevLett.84.1788}{\emph{Phys. Rev. Lett.}} {\bf 84}, 8, 1788 (2000).

\bibitem{okawa2009superconducting} Okawa, M. and Ishizaka, K. and Uchiyama, H. and Tadatomo, H. and Masui, T. and Tajima, S. and Wang, X.-Y. and Chen, C.-T. and Watanabe, S. and Chainani, A. and others. Superconducting electronic state in optimally doped ${Y}$${Ba_2}$${Cu_3}$${O_{7-\delta}}$ observed with laser-excited angle-resolved photoemission spectroscopy. \href{https://doi.org/10.1103/PhysRevB.79.144528}{\emph{Phys. Rev. B}} {\bf 79}, 14, 144528 (2009).

\bibitem{kim2018analysis} Kim, H.-T. Analysis of the diverging effective mass in ${Ya}$${Ba_2}$${Cu_3}$${O_{6+x}}$ for high-${T_c}$ mechanism and pairing symmetry. \href{https://doi.org/10.1142/S0217979218400313}{\emph{Intern. Journal of Modern Physics B}} {\bf 32}, 17, 1840031 (2018).

\bibitem{tsuei2000pairing} Tsuei, C.C. and Kirtley, J.R. Pairing symmetry in cuprate superconductors. \href{https://doi.org/10.1103/RevModPhys.72.969}{\emph{Reviews of Modern Physics}} {\bf 72}, 4, 969 (2000).

\bibitem{welp1994magneto} Welp, U. and Gardiner, T. and Gunter, D. and Fendrich, J. and Crabtree, G.W. and Vlasko-Vlasov, V.K. and Nikitenko, V.I. Magneto-optical study of twin boundary pinning in ${Y}$${Ba_2}$${Cu_3}$${O_{7-\delta}}$. \href{https://doi.org/10.1016/0921-4534(94)91358-7}{\emph{Physica C: Superconductivity}} {\bf 235}, 241-244 (1994).

\bibitem{kim1996paramagnetic} Kim, H.-T. and Minami, H. and Schmidbauer, W. and Hodby, J.W. and Iyo, A. and Iga, F. and Uwe, H. Paramagnetic meissner effect in superconducting single crystals of ${Ba_{1-x}}$${K_x}$${Bi}$${O_3}$. \href{https://doi.org/10.1007/BF00768444}{\emph{Journal of Low Temperature Physics}} {\bf 105}, 3-4, 557-562 (1996). \href{https://arxiv.org/pdf/cond-mat/0206432.pdf}{arxiv.org}.

\bibitem{klemm1997we} Klemm, R.A. Why we still dont know the symmetry of the order parameter in high temperature superconductors. \href{https://pdfs.semanticscholar.org/cab8/63457dff083211059d52cd71bb595e079962.pdf}{\emph Argonne National Lab., IL (United States)} (1997).

\bibitem{tsuei1996pairing} Tsuei, C.-C. and Kirtley, J.R. and Rupp, M. and Sun, J.Z. and Gupta, A. and Ketchen, M.B. and Wang, C.A. and Ren, Z.F. and Wang, J.H. and Bhushan, M. Pairing symmetry in single-layer tetragonal ${Tl_2}$${Ba_2}$${Cu}$${O_{\beta+\delta}}$ superconductors. \href{10.1126/science.271.5247.329}{\emph{Science}} {\bf 271}, 5247, 329-332 (1996).

\bibitem{devereaux1994electronic} Devereaux, T.P. and Einzel, D. and Stadlober, B. and Hackl, R. and Leach, D.H. and Neumeier, J.J. Electronic {Raman} scattering in high-${T_c}$ superconductors: A probe of $d_{x^2-y^2}$ pairing. \href{https://doi.org/10.1103/PhysRevLett.72.396}{\emph{Phys. Rev. Lett.}} {\bf 72}, 3, 396 (1994).

\bibitem{kruchinin2011modern} Kruchinin, S. and Nagao, H. and Aono, S. Modern aspects of superconductivity: theory of superconductivity. \href{https://doi.org/10.1142/7227}{\emph{World Scientific}} (2011).

\bibitem{kruchinin2014physics} Kruchinin, S. Physics of high-${T_c}$ superconductors. \href{https://doi.org/10.1166/rits.2014.1018}{\emph{Reviews in Theoretical Science}} {\bf 2}, 2, 124-145 (2014).

\bibitem{qazilbash2005evolution} Qazilbash, M.M. and Koitzsch, A. and Dennis, B.S. and Gozar, A. and Balci, H. and Kendziora, C.A. and Greene, R.L. and Blumberg, G. Evolution of superconductivity in electron-doped cuprates: Magneto-{Raman} spectroscopy. \href{https://doi.org/10.1103/PhysRevB.72.214510}{\emph{Phys. Rev. B}} {\bf 72}, 21, 214510 (2005).

\bibitem{chen1997electronic} Chen, X.K. and Naeini, J.G. and Hewitt, K.C. and Irwin, J.C. and Liang, R. and Hardy, W.N. Electronic {Raman} scattering in underdoped ${Y}$${Ba_2}$${Cu_3}$${O_{6.5}}$. \href{https://doi.org/10.1103/PhysRevB.56.R513}{\emph{Phys. Rev. B}} {\bf 56}, 2, R513 (1997).

\bibitem{blanc2010loss} Blanc, S. and Gallais, Y. and Cazayous, M. and M{\'e}asson, M.A. and Sacuto, A. and Georges, A. and Wen, J.S. and Xu, Z.J. and Gu, G.D. and Colson, D. Loss of antinodal coherence with a single \textit{d}-wave superconducting gap leads to two energy scales for underdoped cuprate superconductors. \href{https://doi.org/10.1103/PhysRevB.82.144516}{\emph{Phys. Rev. B}} {\bf 82}, 14, 144516 (2010).

\bibitem{hiramachi2007polarization} Hiramachi, T. and Masui, T. and Tajima, S. Polarization dependence of the electronic {Raman} spectra of ${Ca}$-substituted ${YBCO}$ single crystals: as a probe of \textit{s}-wave mixing in the superconducting gap. \href{https://doi.org/10.1016/j.physc.2007.05.009}{\emph{Physica C: Superconductivity and its applications}} {\bf 463}, 89-92 (2007).

\bibitem{zhang2003new} Zhang, Q. and Venturini, F. and Hackl, R. and Hori, J. and Fujita, T. New electronic {Raman} scattering results in underdoped ${La_{2-x}}$${Sr_x}$${Cu}$${O_4}$. \href{https://doi.org/10.1016/S0921-4534(02)02132-9}{\emph{Physica C: Superconductivity and its applications}} {\bf 386}, 282-285 (2003).

\bibitem{kim2017high} Kim, H.-T. High-${T_c}$ mechanism through analysis of diverging effective mass for ${Ya}$${Ba_2}$${Cu_3}$${O_{6+x}}$ and pairing symmetry in cuprate superconductors. \href{10.1142/S0217979218400313}{\emph{arXiv preprint arXiv:1710.07754}} (2017).

\bibitem{klemm2012layered} Klemm, R.A. Layered superconductors. \href{https://books.google.co.kr/books?hl=uk&lr=&id=EWORhjhzdqMC&oi=fnd&pg=PP1&dq=R.+A.+Klemm,+Layered+Superconductors+vol.+1,+(Oxford+Pub.+Co.,+2012)&ots=czRup-9xRv&sig=u7tPtWT_Ha1wHqfhO1iLK5kSrY8&redir_esc=y#v=onepage&q&f=false}{\emph{Oxford University Press}} {\bf 153}, (2012).

\bibitem{yoshida2011pseudogap} Yoshida, T. and Hashimoto, M. and M. Vishik, I. and Shen, Z.-X. and Fujimori, A. Pseudogap, superconducting gap, and {Fermi} arc in high-${T_c}$ cuprates revealed by angle-resolved photoemission spectroscopy. \href{https://doi.org/10.1143/JPSJ.81.011006}{\emph{Journal of the Physical Society of Japan}} {\bf 81}, 1, 011006 (2011).

\bibitem{kaminski2015pairing} Kaminski, A. and Kondo, T. and Takeuchi, T. and Gu, G. Pairing, pseudogap and {Fermi} arcs in cuprates. \href{https://doi.org/10.1080/14786435.2014.906758}{\emph{Philosophical Magazine}} {\bf 95}, 5-6, 453-466 (2015).

\bibitem{razzoli2010fermi} Razzoli, E. and Sassa, Y. and Drachuck, G. and M{\aa}nsson, M. and Keren, A. and Shay, M.i and Berntsen, M.H. and Tjernberg, O. and Radovic, M. and Chang, J. and others. The {Fermi} surface and band folding in ${La_{2-x}}$${Sr_x}$${Cu}$${O_4}$, probed by angle-resolved photoemission. \href{https://doi.org/10.1088/1367-2630/12/12/125003}{\emph{New Journal of Physics}} {\bf 12}, 12, 125003 (2010).

\bibitem{rice2011phenomenological} Rice, T.M. and Yang, K.-Y. and Zhang, F.-C. A phenomenological theory of the anomalous pseudogap phase in underdoped cuprates. \href{https://doi.org/10.1088/0034-4885/75/1/016502}{\emph{Reports on Progress in Physics}} {\bf 75}, 1, 016502 (2011).

\bibitem{shen2005nodal} Shen, K. M and Ronning, F. and Lu, D.H. and Baumberger, F. and Ingle, N.J.C. and Lee, W.S. and Meevasana, W. and Kohsaka, Y. and Azuma, M. and Takano, M. and others. Nodal quasiparticles and antinodal charge ordering in ${Ca_{2-x}}$${Na_x}$${Cu}$${O_2}$${Cl_2}$. \href{10.1126/science.1103627}{\emph{Science}} {\bf 307}, 5711, 901-904 (2005).

\bibitem{yagi2005photoemission} Yagi, H. \href{http://wyvern.phys.s.u-tokyo.ac.jp/f/Research/arch/thesis_yagi.pdf}{\emph{Photoemission study of the high-temperature superconductor ${Y}$${Ba_2}$${Cu_3}$${O_y}$}} (2005).

\bibitem{sassa2011arpes} Sassa, Y. \href{http://doc.rero.ch/record/200236/files/00002351.pdf}{\emph{ARPES investigations on in situ PLD grown $Y$$Ba_2$$Cu_3$$O_{7-x}$}} (2011).

\bibitem{kim2019interplay} Kim, H.-T. Interplay among magnetic, electronic, and superconducting phase transitions in electron-doped cuprate ${Pr_{1-x}}$${La}$${Ce}$${Cu}$${O_{4-x}}$. \href{http://www.kps.or.kr/abstract/view.html?cid=KPS2019B&article_no=931}{\emph{Korean Physical Society Fall Meeting Abstract D5.01}} (2019).

\bibitem{kim2002analyzing} Kim, H.-T. Analyzing Intrinsic Superconducting Gap by Means of Measurement of ${Bi_2}$${Sr_2}$${Ca}$${Cu_2}$${O_{8+x}}$ Superconductors. \href{https://doi.org/10.1143/JPSJ.71.2106}{\emph{Journal of the Physical Society of Japan}} {\bf 71}, 9, 2106-2108 (2002).

\bibitem{di2017disorder} Di Sante, D. and Fratini, S. and Dobrosavljevi{\'c}, V. and Ciuchi, S. Disorder-driven metal-insulator transitions in deformable lattices. \href{https://doi.org/10.1103/PhysRevLett.118.036602}{\emph{Phys. Rev. Lett.}} {\bf 118}, 3, 036602 (2017).

\bibitem{kim2016impurity} Kim, H.-T. Impurity-driven Insulator-to-Metal Transition in {$V$$O_2$}. \href{https://www.jstage.jst.go.jp/article/jpsgaiyo/71.2/0/71.2_1730/_pdf}{\emph{Japan Physical Society Fall Meeting Abstract 13aJB-1}} (2016).

\bibitem{kim2016photoheat} Kim, H.-T. and Kim, M. and Sohn, A. and Slusar, T. and Seo, G. and Cheong, H. and Kim, D.-W. Photoheat-induced {Schottky} nanojunction and indirect {Mott} transition in {$V$$O_2$}: photocurrent analysis. \href{https://doi.org/10.1088/0953-8984/28/8/085602}{\emph{Journal of Physics: Condensed Matter}} {\bf 28}, 8, 085602 (2016).



\end{thebibliography}
\end{document}